\documentclass[10pt,twocolumn]{article}

\usepackage[a4paper,margin=0.75in]{geometry}

\usepackage{titlesec}       
\usepackage{abstract}          
\usepackage{authblk}        

\renewcommand\Affilfont{\small\normalfont\raggedright}
\makeatletter
\renewcommand\AB@affilsepx{\par\protect\Affilfont}
\makeatother

\usepackage{enumitem}
\usepackage{graphicx} 
\usepackage{nicefrac}
\usepackage{amsmath,amssymb,bm} 
\usepackage{siunitx}        
\usepackage{caption}        
\usepackage{booktabs}       
\usepackage[colorlinks=true, linkcolor=blue, citecolor=blue, urlcolor=blue]{hyperref}
\usepackage{fancyhdr}       
\usepackage{titlesec}
\usepackage{setspace}
\usepackage{xspace}
\usepackage{orcidlink}
\usepackage[dvipsnames]{xcolor}
\usepackage[backend=biber, sorting=none, style=nature]{biblatex}
\title{Massively parallel transdimensional sampling with data assimilation:\\
an application to LISA galactic binaries}

\author[1,2]{Gabriele Demasi\textsuperscript{$\star$}\orcidlink{0009-0009-5320-502X}}
\author[3, 4]{Walter Del Pozzo
    \orcidlink{0000-0003-3978-2030}}

\affil[1]{Dipartimento di Fisica e Astronomia, Università degli Studi di Firenze,
Via Sansone 1, Sesto Fiorentino (Firenze) I-50019, Italy}
\affil[2]{
INFN, Sezione di Firenze, Sesto Fiorentino (Firenze) I-50019, Italy}
\affil[3]{Dipartimento di Fisica “Enrico Fermi”, Università di Pisa,
Largo Bruno Pontecorvo 3, Pisa I-56127, Italy}
\affil[4]{INFN, Sezione di Pisa, Largo Bruno Pontecorvo 3, Pisa I-56127, Italy}

\date{}
\begin{document}

\twocolumn[
\maketitle
\begin{onecolabstract}
\noindent
  Transdimensional Bayesian inference is becoming a key ingredient in
  gravitational-wave astronomy. In many relevant applications, the number of
  components needed to describe the data is not known a priori and must be
  inferred jointly with their parameters. 
  We present a transdimensional Sequential Monte Carlo (SMC) framework designed to exploit massive parallelism. The method evolves a population of particles through a sequence of tempered distributions and uses the  No-U-Turn Sampler for the exploration of fixed-dimensional parameter space while reversible-jump birth and death moves allow the number of components to vary. The same SMC construction also enables posterior samples obtained from a shorter data segment to be updated as additional data become available, enhancing the overall efficiency of the analysis. We validate the approach on two proof-of-concept problems: the recovery of a sequence of Gaussian pulses and a simplified LISA Galactic-binary inference problem.
  In both cases, the method recovers the injected number of components and produces consistent posterior estimates. 
   These results indicate that the proposed method is a promising avenue toward scalable and parallel Bayesian transdimensional inference for current and future gravitational-wave astronomy.
\end{onecolabstract}
]
\vspace{1em}

\footnotetext{%
\textsuperscript{$\star$}~E-mail: \texttt{gabriele.demasi@unifi.it}%
}


\section{Introduction}\label{sec:introduction}
\noindent
Bayesian inference is widely used in gravitational-wave astronomy because it provides a principled framework for updating prior information about astrophysical hypotheses in light of observed data.
\cite{ThraneTalbot:2019}. 

For compact-binary coalescences 
observed by the LIGO-Virgo-KAGRA \cite{2015:LIGO, 2015:VIRGO, 2013:KAGRA} interferometers, such as  binary black holes (BBHs), Bayesian inference is used to infer
the masses, spins, distance, sky location, inclination, and other source
properties from the observed strain data \cite{Veitch:2015lalinference,Ashton:2019bilby}, 
in a process called parameter estimation. These examples are a typical scenario that requires "fixed-dimensional" inference, 
i.e. the number of parameters describing the model is known and fixed a priori. \\
Because gravitational-wave parameter spaces are often high dimensional, Bayesian inference is typically performed using stochastic samplers such as Markov Chain Monte Carlo \cite{Speagle:2019_MCMC} or nested sampling\cite{Ashton:2019bilby}. More recently, approaches based partly or entirely on machine learning have also become increasingly common \cite{2021:Williams_Nessai, 2023:Wong_JIM,Dax:2021dingo}.
Recently, Sequential Monte Carlo methods have also been shown to provide a powerful alternative for gravitational-wave inference.\cite{2025:Williams_Validating_SMC,Williams:2025aar,  Demasi:2026ltw}\\
As the detector sensitivity increases and new instruments come online, the inference problem is severely complicated. For instance, future detectors such as Einstein Telescope \cite{Punturo_2010} and LISA (Laser Interferometer Space Antenna)\cite{AmaroSeoane:2017lisa} will observe millions of sources, implying that any given data segment will contain an unknown number of sources with unknown parameters. 
It is therefore imperative to construct inference frameworks that allow for parameter spaces whose dimensionality is also inferred from the data. Already current detectors require such frameworks. For instance, the number of instrumental or phenomenological features can vary in noise modelling \cite{2014:Bayeswave_cornish}; while mixture models or subpopulations may be introduced in astrophysical population inference. \cite{2023:Toubiana_RJCMCM}.
The most challenging task however, is the analysis of LISA data. LISA will be a space-based GW observatory sensitive to signals in the mHz regime. Its data stream will
contain thousands of overlapping signals, including Galactic binaries, massive
black-hole binaries, extreme-mass-ratio inspirals, stochastic backgrounds, and
instrument noise \cite{AmaroSeoane:2017lisa}. It is now widely recognized that all these sources, despite their exact number being unknown, must be inferred simultaneously through a procedure commonly referred to as the \textit{Global Fit}.

The LISA Global Fit is therefore
intrinsically trans-dimensional: the number of sources in the data is part of
the inference problem.

The standard approach for dealing with an unknown dimensionality of the parameter
space is Reversible Jump Markov Chain Monte Carlo (RJMCMC)
\cite{Green:1995rj}.  RJMCMC explores not only the continuous parameters of a
model, but also a variable number of dimensions through moves that connect
models with different parameter counts. 
This idea is a key ingredient in
existing LISA global-fit approaches,
\cite{Littenberg:2020galactic,Littenberg:2023prototype,Lackeos:2023radler,Strub:2024global, Katz:2024oqg}, but also in astrophysical binary black hole population and instrument noise modeling.
At the same time, alternative strategies for trans-dimensional sampling have recently begun to appear, both based 
on stochastic algorithms \cite{Astorino:2025ccl} and/or machine learning \cite{Houba:2025dnr}.

Despite their success, traditional trans-dimensional inference strategies face
important limitations in the regimes targeted here. 
 First, MCMC trajectories
are intrinsically sequential, whereas future GW analyses, in particular
in the LISA context, will benefit from algorithms that expose a high degree of
parallelism. 
 Second, those algorithms are often based on random-walk moves, that explore high-dimensional
fixed-model parameter spaces inefficiently, especially when the posterior is highly multi-modal 
and strongly correlated. 

 Third, when new data become available,
as naturally happens for long-duration signals or for the progressive
accumulation of LISA observations, an analysis is often restarted rather than
updated from the previous posterior.  This is particularly relevant for
low-latency applications and for LISA sources that remain in band for months or
years.

Sequential Monte Carlo provides a natural way to address several of these
issues.  SMC evolves a population of particles from a reference distribution,
usually the prior, to the posterior through a sequence of tempered
distributions \cite{DelMoral:2006smc,Chopin:2020smc}. 
Because the particles
are mutated independently, SMC is naturally
parallel and can naturally exploit modern hardware accelerators.  Moreover, the
temperature ladder provides an adaptive bridge from a broad prior to a
concentrated posterior and yields an estimate of the Bayesian evidence. 
In this work we address the limitations above by embedding (potentially generic) trans-dimensional moves inside an SMC scheme,
as already suggested, for instance, in \cite{radial_basis_SMC, transdimensional_data_ass}.
We use a gradient-based method, the No-U-Turn Sampler, for the
  fixed-model mutations,  building on the work in \cite{Demasi:2026sharpy}, that
showed that combining SMC with the No-U-Turn Sampler (NUTS) provides a fast and
accurate likelihood-based sampler for fixed-dimensional GW inference.
Additionally, we exploit the SMC scheme to update posterior samples as new data become available. 

We validate this approach as a proof of concept on two problems. The first is a toy model, in which a sequence of Gaussian pulses is injected\footnote{We will consistently use the technical jargon ``to inject'' to describe the process of simulating a signal and adding it to the noise} in the data and subsequently recovered.
The second is a highly idealised LISA global-fit-like problem where we attempt to recover galactic binaries injected into the data. 

The rest of the paper is
organized as follows.  In
sec. \ref{sec:SMC} we review the Sequential Monte Carlo algorithm;  in
sec. \ref{sec:transition_kernel} we describe how we explore the trans-dimensional space within the SMC scheme, briefly describing the RJMCMC and NUTS; in sec. \ref{sec:implementation} we outline  the algorithm and we  discuss some practical implementation details and  finally, we draw our conclusions in
sec.~\ref{sec:Conclusions}.

\section{Sequential Monte Carlo}\label{sec:SMC}
Sequential Monte Carlo is an algorithm that uses a sequence of intermediate distributions to bridge between an initial distribution and a final one. 
For static problems, where the amount of data in the likelihood is fixed, the initial distribution is often taken to be the prior distribution while the posterior acts as a target. 
With this in mind, an inverse temperature $\beta_t$ is introduced and the bridge distribution   $p(\boldsymbol{\Theta}| \beta)$
can be defined as follows:
\begin{equation}
  \label{eq: prior_bridge}
    p(\boldsymbol{\Theta}| \beta) = 
\frac{\mathcal{L}(d|\boldsymbol{\Theta})^{\beta_t}
 \mathcal{\pi}(\boldsymbol{\Theta})}{\mathcal{Z}_t}, 
\end{equation}
such that when $\beta_t = 0 ( \beta_t=1)$ we recover the prior (posterior).

Let us look instead at the situation where we have already observed a certain amount of data $d_1$ that led to a posterior  $ p(\boldsymbol{\Theta}_1|d_1)$. 
When we observe a new stretch of data such that the total amount of data is now $d_2$, we can define a bridge such that the initial distribution is $p(\boldsymbol{\Theta}|d) $ while the target is $p(\boldsymbol{\Theta}|d_2) $:
\begin{equation}
  \label{eq: reuse_bridge}
    p(\boldsymbol{\Theta}| \beta)  \propto p(\boldsymbol{\Theta}|d_1)  \left(\frac{p(\boldsymbol{\Theta}|d_2)}{p(\boldsymbol{\Theta}|d_1)}\right)^\beta.
\end{equation}

The SMC starts with $\beta_0=0$,
and finishes when $\beta_T=1$.  Each SMC iteration is made of three
steps:

\begin{enumerate}
  \item \underline{\textbf{Reweighting}}\\
  At iteration $t-1$, each particle is assigned an incremental weight that
  measures how compatible it is with the next tempered target:
  \begin{equation}
    w_t^{(i)} = \frac{p(\boldsymbol{\Theta}_{t-1}^{(i)}\mid\beta_{t})}
         {p(\boldsymbol{\Theta}_{t-1}^{(i)}\mid\beta_{t-1})}.
    \label{eq:incremental_weight}
  \end{equation}
  If two consecutive distributions are too far apart, the weights become
  uneven and the particle approximation is dominated by a small number of
  points. This effect is monitered through the effective sample size:
  \begin{equation}
    \mathrm{ESS}_t =
    \frac{
      \left(\sum_{i=1}^{N_\mathrm{P}} w_t^{(i)}\right)^2
    }{
      \sum_{i=1}^{N_\mathrm{P}} \left(w_t^{(i)}\right)^2
    }.
    \label{eq:ess}
  \end{equation}
  The temperature is chosen adaptively by solving
  \begin{equation}
    \mathrm{ESS}(\beta_t)-\alpha N_\mathrm{P}=0,
    \label{eq:ess_target}
  \end{equation}
  where $\alpha \in (0,1]$ controls the fraction of effective particles retained
  at each step, providing an automatic strategy for annealing, that dramatically helps the exploration of the parameter space in case of narrow and highly multimodal parameter spaces.  In practice, Eq.~\eqref{eq:ess_target} is solved by bisection
  with the constraint $\beta_t\leq 1$.
  In the reweighting step, the same incremental weights also provide an estimate of the normalizing
constant $\mathcal{Z}$. 
The ratio between two consecutive normalizing constants at step $t$ is estimated as\cite{Chopin:2020smc}:
\begin{equation}
            \frac{\mathcal{Z}_t}{\mathcal{Z}_{t-1}} = \frac{1}{N}\sum_{i= 1}^{N}  w^{(i)}_t.
\end{equation}
If the prior is properly normalized, and hence $Z_0 =1$, the final evidence after $T$ iterations is computed as:
\begin{equation}
            \mathcal{Z} =  
            \prod_{t=1}^{T}  \frac{\mathcal{Z}_t}{\mathcal{Z}_{t-1}}.
\end{equation}

  \item \underline{\textbf{Resampling}}\\
  The particles are resampled according to their normalized weights.  This
  replaces low-weight particles, typically lying in regions that are becoming
  disfavored by the likelihood, with copies of particles in higher-likelihood
  regions.  As $\beta_t$ increases, the likelihood progressively sharpens the
  target distribution and the resampling step concentrates computational effort
  where the posterior mass is building up.

  \item \underline{\textbf{Mutation}}\\
  Finally, the resampled particles are propagated with a transition kernel $\mathcal{K}$.
  This step prevents particle degeneracy
  and restores diversity after resampling. 
  
  The mutation part is the most computationally expensive one but, since the particles are mutated
  independently, massive parallelization is possible.

  The efficiency of the mutation kernel determines how well the particle cloud follows the tempered
  targets and therefore how many SMC iterations are needed to reach
  $\beta=1$.

\end{enumerate}

\section{Transition kernel}\label{sec:transition_kernel}
The Sequential Monte Carlo scheme is general and agnostic about the type of transition kernel employed in the mutation part, as long as it leaves the target distribution invariant. 
The efficiency of this kernel is crucial: it determines how well the particle population explores the target distribution and how many SMC iterations are required to reach the final posterior.

Gradient-based methods, such as the  Hamiltonian Monte Carlo \cite{Neal:2011hmc}, and its extension the No-U-Turn-Sampler\cite{Hoffman:2014nuts}, have a higher sampling efficiency than traditional random-walk-based samplers, since they use the  gradient information of the posterior to explore the parameter space. 
However, their use is limited to fixed-dimensional problems, since it is non trivial to adapt the Hamiltonian dynamics in a space with a variable number of dimensions.

We adopt here a hybrid approach such that the mutation is performed by a mixture of transition kernels. One component of the mixture is $K^{\mathrm{NUTS}}$ responsible for updating the parameters in the fixed dimensional case, with some probability $ p_{\mathrm{NUTS}}$.  For the transdimensional moves we rely on a RJMCMC kernel that proposes to add a new component $K^{\mathrm{RJ}, \mathrm{birth}}$ with some probability $ p_{\mathrm{birth}}$, and a death kernel $ K^{\mathrm{RJ}, \mathrm{death}}$ that, with some probability $ p_{\mathrm{death}}$ proposes to remove a component from the data.

Therefore, the mutation kernel $\mathcal{K}$ of the SMC is: 

\begin{equation}\label{eq:kernel_mixture}
    \mathcal{K} = 
    p_{\mathrm{NUTS}} K^{\mathrm{NUTS}}
    +
    p_{\mathrm{birth}} K^{\mathrm{RJ}, \mathrm{birth}}
    +
    p_{\mathrm{death}} K^{\mathrm{RJ}, \mathrm{death}} ,
\end{equation}

\subsection{Reversible Jump Markov Chain Monte Carlo }
Let $k$ denote the model index which is indicates, in our case, the number of components in the data.

The full sample space is:

\begin{equation}
    \boldsymbol{\Theta}
    =
    \bigsqcup_{k=0}^{K_{\max}}
    \left(\{k\}\times\theta_k\right), 
  \end{equation}
where $\theta_k$ is the set of continuous parameters that describe each component.

The posterior is then given by:
\begin{equation}
    p(k,\boldsymbol{\theta}|\boldsymbol{d})
=
\frac{
p(k)\,p(\boldsymbol{\theta}\mid k)\,p(\mathbf{d}\mid \boldsymbol{\theta},k)
}{
\displaystyle
\sum_{k\in\mathcal{K}}
\int
p(k)\,p(\boldsymbol{\theta}\mid k)\,p(\mathbf{d}\mid \boldsymbol{\theta},k)
\,d\boldsymbol{\theta}
}
\end{equation}

In a transdimensional problem, we wish to sample both the continuous
parameters $\boldsymbol{\theta}_k$ and the model index $k$. Reversible Jump
Markov Chain Monte Carlo (RJMCMC) extends the Metropolis--Hastings algorithm to
this setting by allowing proposals between parameter spaces of different
dimension \cite{Green:1995rj}.  A move from
$(k,\boldsymbol{\theta}_k)$ to $(k',\boldsymbol{\theta}'_{k'})$ is constructed
by drawing auxiliary variables $\boldsymbol{u}$ and applying an invertible map
\begin{equation}
  (\boldsymbol{\theta}'_{k'},\boldsymbol{u}')
  =
  T_{k\rightarrow k'}(\boldsymbol{\theta}_k,\boldsymbol{u}),
\end{equation}
with the dimension-matching condition
\begin{equation}
  \dim(\boldsymbol{\theta}_k)+\dim(\boldsymbol{u})
  =
  \dim(\boldsymbol{\theta}'_{k'})+\dim(\boldsymbol{u}') .
\end{equation}
The proposal is accepted with probability
\begin{equation}
  \alpha =
  \min\left[
  1,
  \frac{
    p(k',\boldsymbol{\theta}'_{k'})
    r_{k'\rightarrow k}
    q_{k'\rightarrow k}(\boldsymbol{u}'\mid k',\boldsymbol{\theta}'_{k'})
  }{
    p(k,\boldsymbol{\theta}_k)
    r_{k\rightarrow k'}
    q_{k\rightarrow k'}(\boldsymbol{u}\mid k,\boldsymbol{\theta}_k)
  }
  \left|\mathcal{J}
  \right|
  \right],
  \label{eq:rj_acceptance}
\end{equation}
where $r_{k\rightarrow k'}$ is the probability of proposing the move type,
$q_{k\rightarrow k'}$ is the proposal density for the auxiliary variables, and
the final term  $   \left|\mathcal{J}
\right| =
\frac{\partial(\boldsymbol{\theta}'_{k'},\boldsymbol{u}')}
     {\partial(\boldsymbol{\theta}_k,\boldsymbol{u})}$ 
  is the Jacobian determinant of the transformation.

In this work we consider the simple nested case in which neighboring models are
connected by adding or removing one component, $k'=k\pm1$, so that $r_{k\rightarrow k+1} = p_{\text{birth}}$ and $r_{k\rightarrow k-1} = p_{\text{death}}$.
For this append/remove map the Jacobian is unity, so Eq.~\eqref{eq:rj_acceptance}
reduces to the usual Metropolis--Hastings ratio with the appropriate forward and
reverse proposal probabilities.  In the baseline implementation,
$\boldsymbol{\vartheta}$ is proposed from the prior for a single component, while
the reverse move removes one of the existing components.

\subsection{No-U-Turn-Sampler}\label{sec:NUTS}

As highlighted in the previous section, a good transition kernel is fundamental
to maintain a high ESS while using a small number of SMC iterations.  Standard
MCMC kernels typically propose stochastic jumps in parameter space and can
therefore suffer from random-walk behavior. 

Gradient-based sampling methods enhance the efficiency of the exploration of the parameter space by using the gradient information of the posterior to partially suppress the random walk behavior. 
The Hamiltonian Monte Carlo does so by augmenting the parameter space of interest  $\boldsymbol{\theta}$ with random auxiliary momentum variables $\boldsymbol{r}$ drawn from a multivariate normal distribution:
\begin{equation}
    \boldsymbol{r} \sim \mathcal{N}(0, \mathcal{M}), 
\end{equation}
where $\mathcal{M}$ is the so-called mass matrix, which can also be interpreted as a metric in the parameter space. 

Then, using the log-posterior as potential energy and a kinetic energy using the sampled momentum, a Hamiltonian is introduced:

\begin{equation}
  \mathcal{H}(\boldsymbol{\theta}, \boldsymbol{\mathrm{r}}) = \  U(\boldsymbol{\theta})+ K(\boldsymbol{\mathrm{r}}) = -\log p(\boldsymbol{\theta|d)} + \dfrac{1}{2}\boldsymbol{r}^T\mathcal{M}^{-1}\boldsymbol{r}
\end{equation}

To propose a new point, the Hamiltonian system is evolved using Hamilton's equations:
\begin{align}
  \frac{d\boldsymbol{\theta}}{dt} &= \frac{\partial H}{\partial \mathbf{r}} = \nabla_{\boldsymbol{r}}K(\boldsymbol{{r}}) \,,\\
  \frac{d\mathbf{r}}{dt} &= -\frac{\partial H}{\partial \boldsymbol{\theta}} = -\nabla_{\boldsymbol{\theta}} U(\boldsymbol{\theta})\,.
\end{align}
The proposed point is then accepted or rejected based on the difference between the initial and final values of the Hamiltonian.

In  practice, Hamilton's equations are solved numerically with a symplectic integrator.  The integration
time is one of the main tuning parameters of HMC: too short a trajectory
recovers random-walk-like behavior, while too long a trajectory can waste
computation by returning close to the starting point.

The No-U-Turn Sampler addresses this issue by adapting the trajectory length
during the proposal.  It stops the integration when the trajectory starts
turning back toward its initial point, producing an automatically tuned HMC
transition without fixing the number of leapfrog steps in advance
\cite{Hoffman:2014nuts}.  

Additionally, also the mass matrix is a non-trivial parameter to tune. Since it can be seen as the metric of the parameter space, we follow the strategy employed in \cite{Demasi:2026sharpy}, setting it to the Hessian of the negative log posterior.

\section{Algorithm and implementation}\label{sec:implementation}

One SMC run proceeds as follows: 
\begin{enumerate} 
  
  \item If in the \texttt{prior-start} mode, draw $N$ particles from $p(k, \theta_k)$ and set $\beta_0=0$. If in the \texttt{posterior-start} mode, use particles obtained in a previous run.
  
  \item Choose the next $\beta_t$ using the ESS condition in Eq.~\eqref{eq:ess}.
  
  \item Reweight the particles using Eq.~\eqref{eq:incremental_weight}.
  \item Resample.
  \item Mutate each particle with the kernel in Eq. ~\eqref{eq:kernel_mixture}, using eq. \eqref{eq: prior_bridge} if in the \texttt{prior-start} case or \eqref{eq: reuse_bridge} in the \texttt{posterior-start} case.
  
  \item Repeat until $\beta_t=1$.

\end{enumerate}

The SMC scheme offers a clear advantage for parallel implementation as each particle at the mutation step is completely independent of the others. This makes the scheme naturally suited for modern hardware accelerators such as GPUs and TPUs.

We exploit the intrinsic parallelism of the algorithm by implementing it in \texttt{JAX}, a Python library for high-performance computing whose primary features include the automatic differentiation for efficiently computing gradients, automatic vectorization and Just-In-time compilation, while being device agnostic, meaning that the same code can be run on CPUs or GPUs.
For the NUTS kernel we rely on the BlackJAX\cite{BlackJAX:2024} implementation. 

A practical implementation detail is that particles must be grouped according
to their current model dimension $k$ before applying the within-model mutation
kernel.
The reason is
that \texttt{JAX} just-in-time compilation requires fixed array shapes, whereas
the dimension of $\theta_k$ changes with $k$. Particles with different values of
$k$ therefore cannot be passed to a single batched NUTS kernel. Instead, we
partition the population into fixed-dimensional groups and apply the compiled
NUTS transition separately to each group. This retains the benefits of batching
and vectorization within each model.

\section{Applications}
We test the method on two problems in which the number of components in the data is unknown: (i) a toy model containing a sequence of Gaussian pulses and (ii) a simplified LISA Galactic-binary inference problem.

In both cases we consider two approaches for the inference:
\begin{itemize}

 \item In the \texttt{prior-start} approach, each data segment is analyzed from scratch, without using information from analyses of shorter data segments. In practice, the SMC evolution starts from the prior and uses the bridge distribution in Eq.~\eqref{eq: prior_bridge}.

  \item In the \texttt{posterior-start} approach, progressively longer data segments are analyzed by reusing posterior samples obtained from shorter segments. In practice, the SMC evolution uses the bridge distribution in Eq.~\eqref{eq: reuse_bridge}.

\end{itemize}

\subsection{Sequence of gaussian pulses}\label{sec:toy_validation}
In this example we consider a sequence of 10 Gaussian pulses in 200 seconds of data in the absence of noise, to illustrate the main features of the algorithm without the complication of realistic scenarios.

\begin{figure}
  \centering
  \resizebox{0.5\textwidth}{!}{
\includegraphics{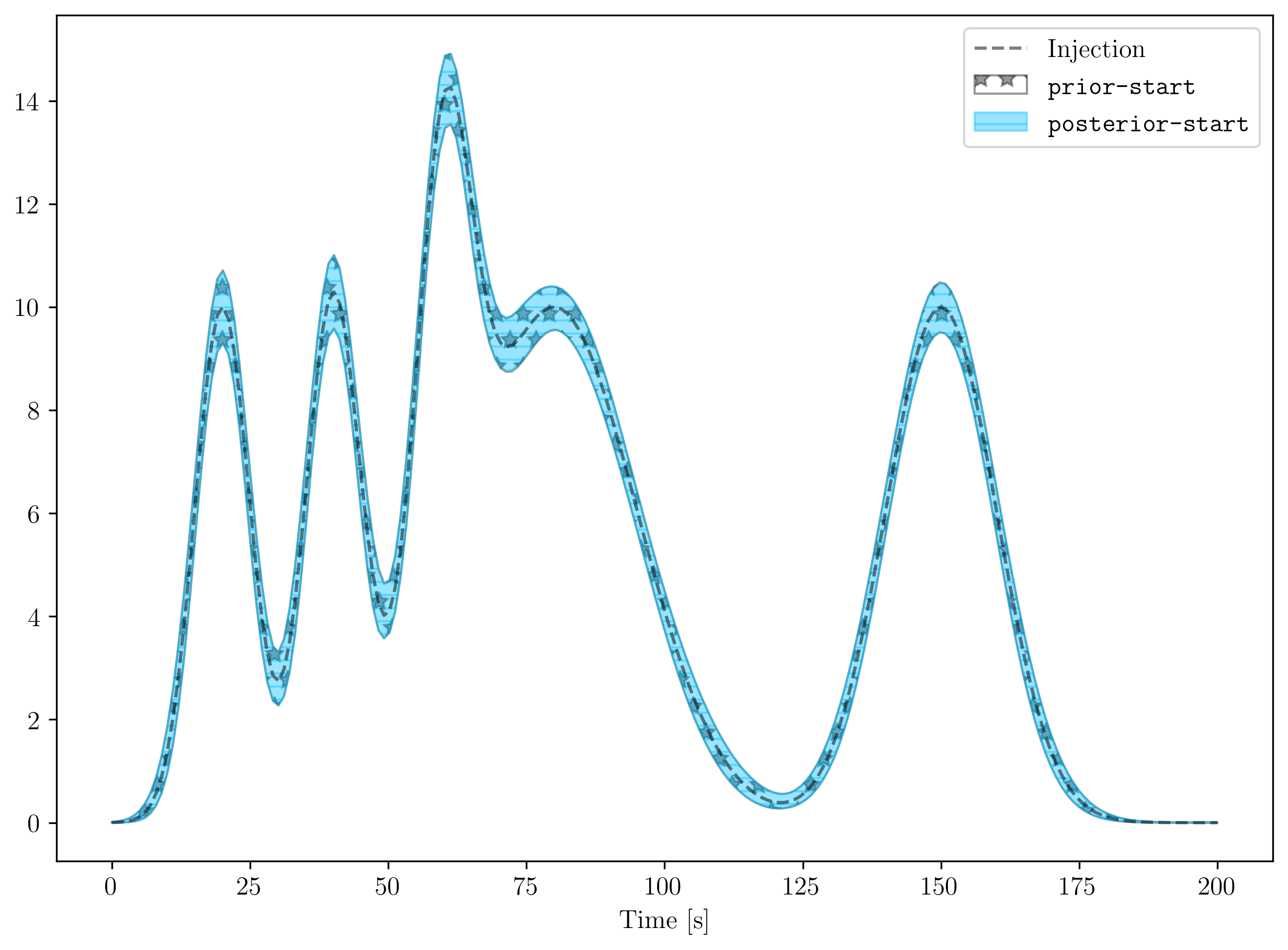
}}
  \caption{Sequence of Gaussian pulses described in Sec.~\ref{sec:toy_validation}. The black dashed line shows the injected signal as a function of time. The two differently hatched shaded regions, that are practically indistinguishable,  show the 90\% credible intervals for the inferred signal in the \texttt{posterior-start} and \texttt{prior-start} analyses.} 
  \label{fig:predictive_gaussian}
\end{figure}

\begin{figure}
  \centering
  \resizebox{0.5\textwidth}{!}{
\includegraphics{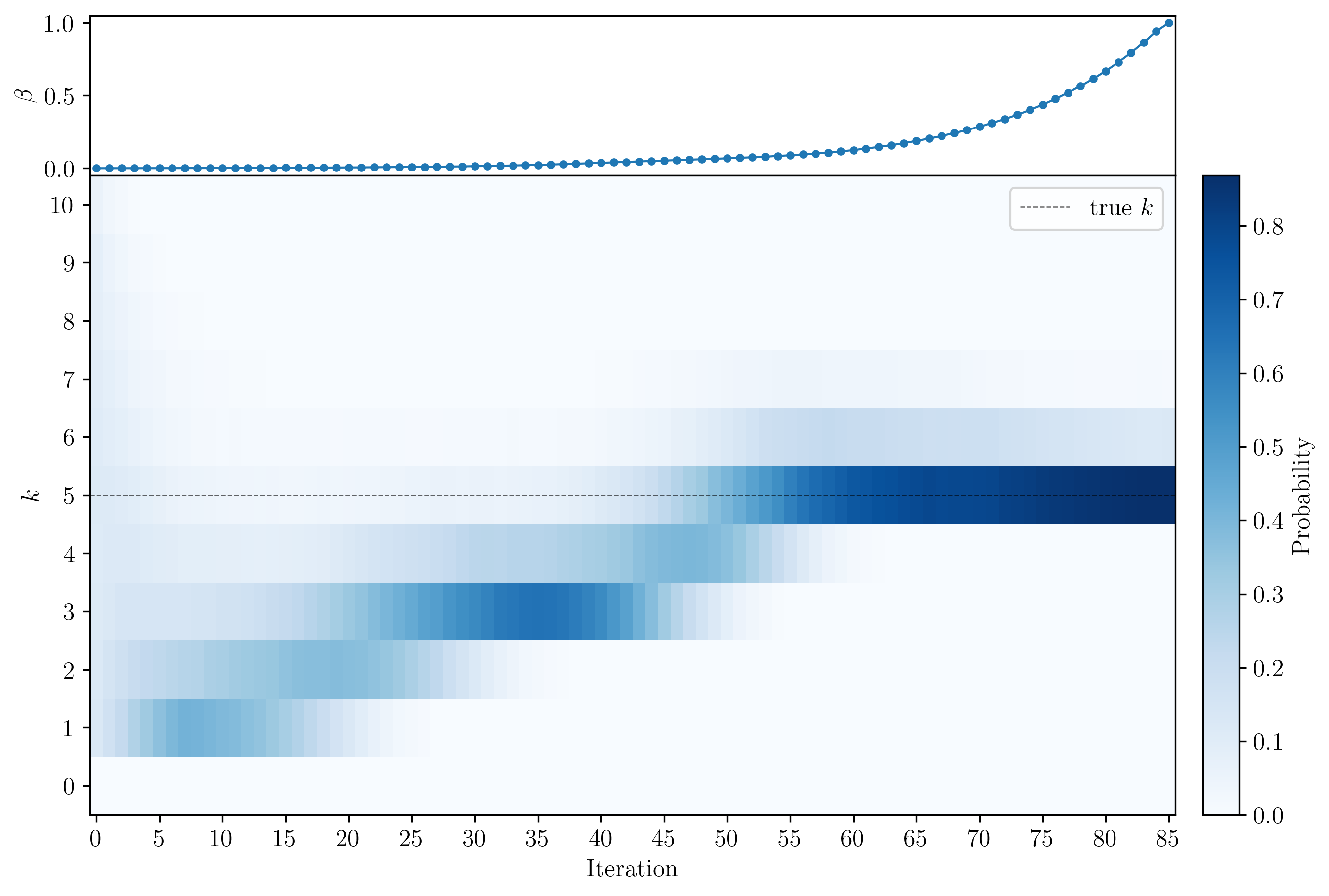
}}
  \caption{
    \textit{Bottom panel:}
   Evolution of the posterior over $k$ as a function of the SMC iterations for the gaussian pulse example (sec. \ref{sec:toy_validation}). For each iteration, we report the histogram of the number of component $k$ as a heatmap, such that darker colors indicate higher posterior probability.  
  \textit{Top panel:}
  Behaviour of the inverse temperature $\beta$ as a function of the SMC iterations.
  }
  \label{fig:heatmap_gaussian}
\end{figure}

Each Gaussian pulse is described with the following parameters $\theta_j  = \{ A, \mu, \sigma\}$, that are respectively the amplitude, the mean and the variance of the pulse. The prior distribution is taken to be uniform in each of the parameters.
The injected signal can be seen in fig. \ref{fig:predictive_gaussian}, as a dashed line. 
We run the analysis in the time domain.
We first apply the method to the whole 200 seconds of data, in the \texttt{prior-start} case. We use 6000 particles per iteration, a constant fraction of effective samples $\alpha = 0.9$ (see equation \eqref{eq:ess_target}), uniform priors on the parameters, acting also as birth proposal, a uniform death proposal,  and the probability to perform a fixed-dimensional mutation with NUTS, $p_{\text{NUTS}}$, to do a birth $p_{\text{birth}}$ or a death $p_{\text{death}}$ respectively to 0.6, 0.2 and 0.2.

The top panel of Fig.~\ref{fig:heatmap_gaussian} shows the evolution of the inverse temperature $\beta$, while the bottom panel shows the posterior distribution over k across SMC iterations. 
We observe that in the very first stages of the SMC, because the bridge distribution is close to the prior, the probability for each possible number of components is uniformly distributed between the allowed prior range, that is [1,10] in this case. At intermediate temperatures, the posterior over the number of components peaks below the injected value. This is expected because the tempered likelihood still favors simpler models. In the final SMC iterations, as $\beta$ approaches one, the posterior over $k$ peaks at the injected value.
In fig. \ref{fig:predictive_gaussian} we plot the $90\%$ confidence interval posterior distribution of the recovered signal as shaded region with vertical hatches, showing an excellent agreement between the posterior estimate and the injected value.

\begin{figure}
  \centering
  \resizebox{0.5\textwidth}{!}{
\includegraphics{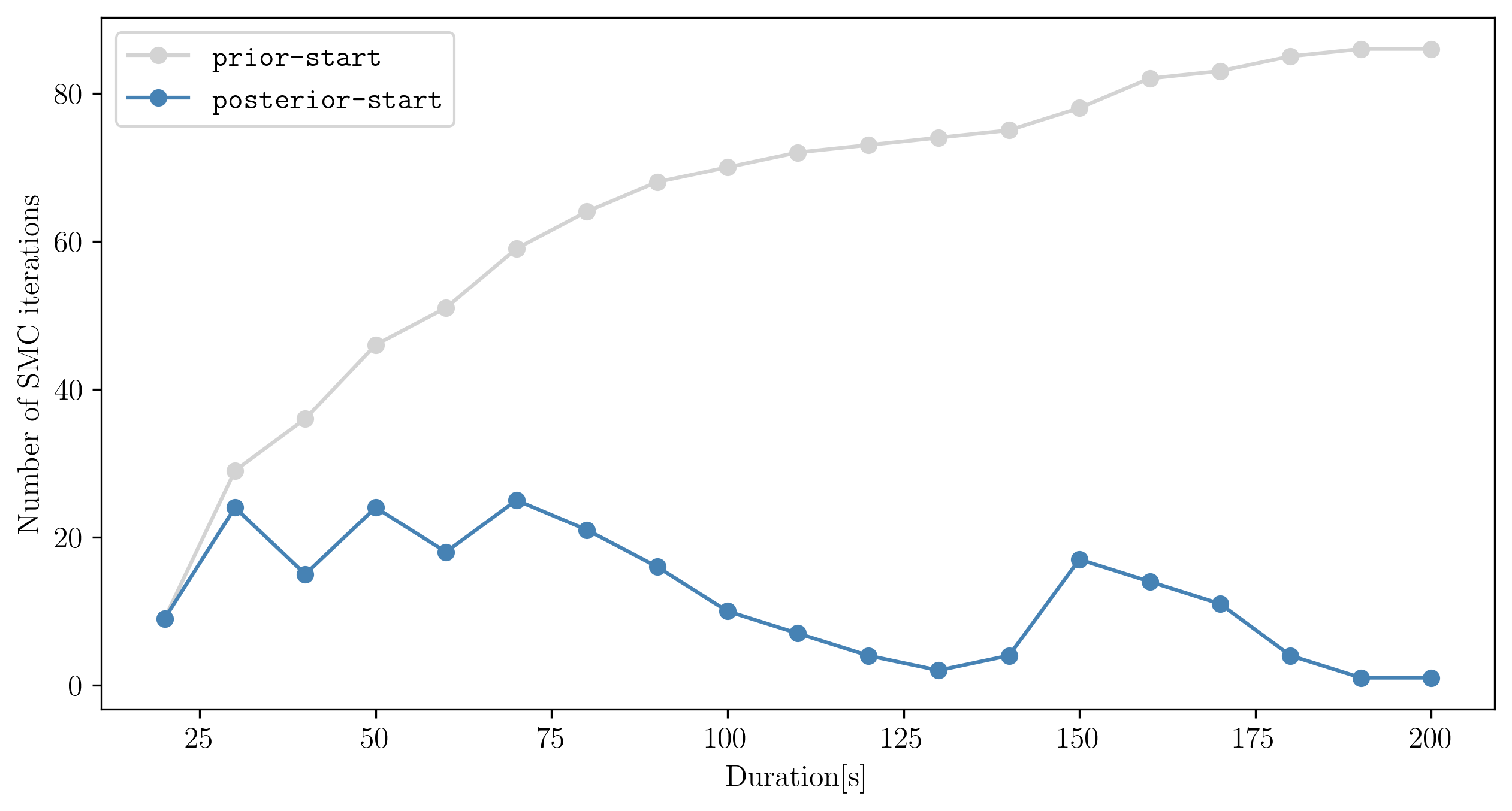
}}
  \caption{Comparison between the total number of SMC iterations in the \texttt{prior-start} and \texttt{posterior-start}  case as a function of the duration of the portion of the signal considered, for the gaussian pulses case (sec. \ref{sec:toy_validation}). }
  \label{fig:iterations_comparison_gaussian}
\end{figure}

\begin{figure}[h!]
  \centering
  \resizebox{0.5\textwidth}{!}{
\includegraphics{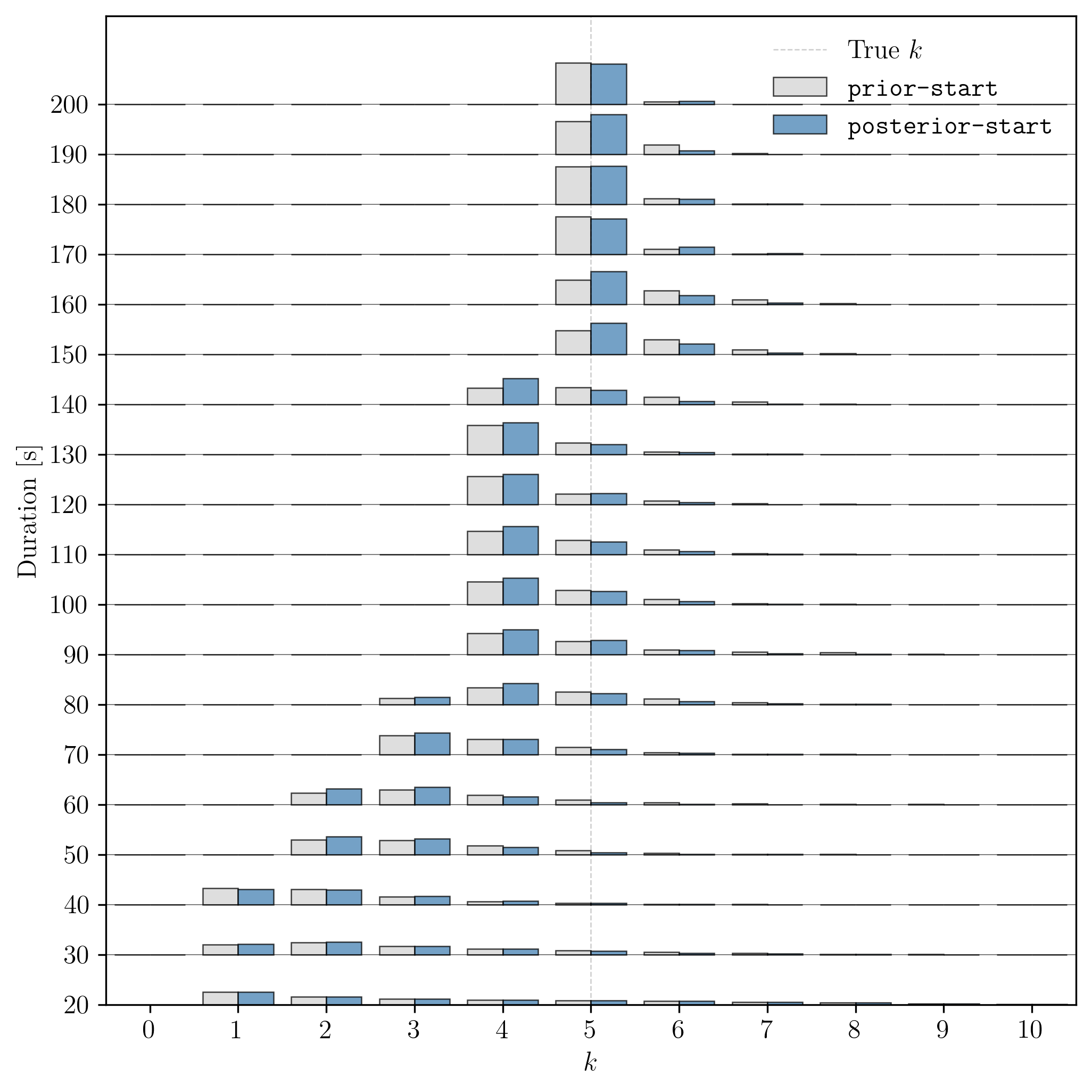
}}
  \caption{Comparison between the posterior over the number of  gaussian pulses found in data for the \texttt{prior-start} and \texttt{posterior-start}  case as a function of the duration of the portion of data considered. }
  \label{fig:ridgeline_gaussians}
\end{figure}

We now consider the \texttt{posterior-start} case. 
We consider segments of data with a progressively larger duration, such that the subsequent segment contains the segment before while being 20 seconds longer. The duration of each segments therefore varies from 20 seconds to 200 seconds. For the segments with duration greater than $20 s $we start the analysis using the results already obtained with the shorter segment, without the need of running the whole analysis from scratch. 
For comparison, we also run the analysis  for each segment in the 
\texttt{prior-start} mode.

Fig. \ref{fig:iterations_comparison_gaussian}  shows the number of SMC iterations needed to converge from the base to the target distribution in the two cases considered, as a function of the total duration of the signals.  As expected, this number increases in the \texttt{prior-start} case while it decreases in the \texttt{posterior-start} case, since the SMC starts from a more similar distribution and therefore fewer tempering steps are needed. It may also be worth noting the bump in the \texttt{posterior-start} case at approximately 150 seconds. This behavior can be understood from the morphology of the signal: a new pulse appears at approximately this time, requiring additional SMC iterations to explore the new region of parameter space.
This is a typical behavior that has been observed in our experiments.

Additionally in fig.  \ref{fig:iterations_comparison_gaussian}, we report a comparison between the posteriors on the number of Gaussian pulses present in data found with in the two cases considered, as a function of the duration of the signal, showing a good agreement. Finally, in fig \ref{fig:predictive_gaussian}, we show the $90 \%$ credible interval on the posterior in the \texttt{posterior-start} case as a shaded region with horizontal hashed, identical to the one obtained in the other case considered.

\subsection{LISA Galactic binaries}
\label{sec:lisa_application}

\begin{figure}[h!]
  \centering
  \resizebox{0.5\textwidth}{!}{
\includegraphics{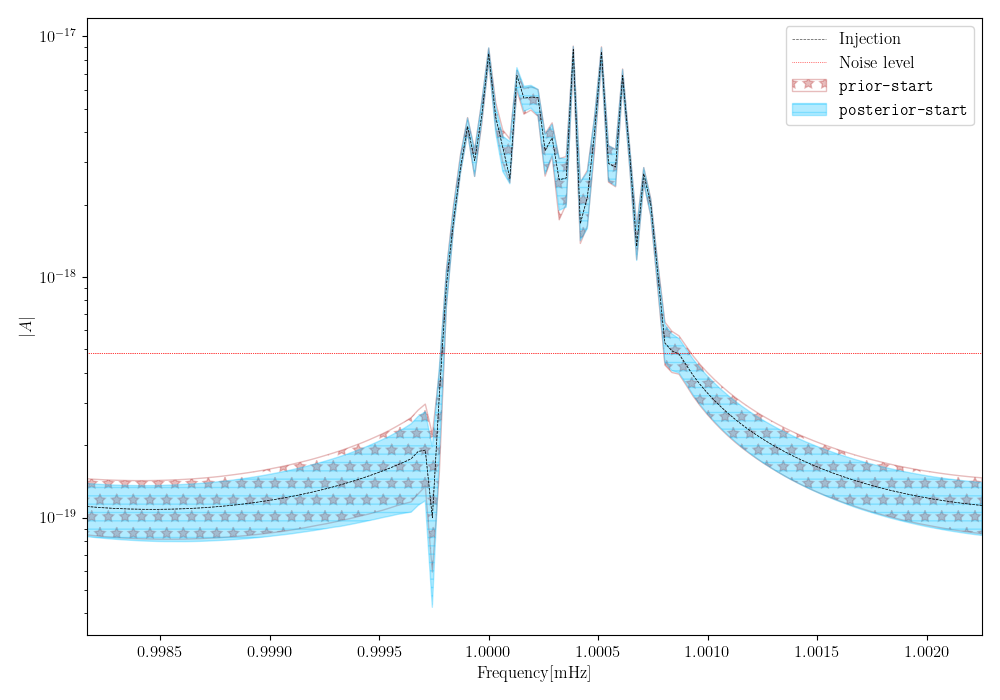
}}
  \caption{Simulated data resulting from the superposition of 10 galactic binaries, as described in section \ref{sec:lisa_application}. The dashed black line represents the injected value while the shaded region with horizontal(vertical) hashed represent the $90\%$ confidence region on the inferred signal in the \texttt{posterior-start} (\texttt{prior start}) case. }
  \label{fig:signal_recovery_lisa}
\end{figure}

Space-based gravitational-wave interferometers such as LISA will observe data streams containing thousands of overlapping signals. It is now well established that LISA data analysis should be carried out within a global-fit framework, in which all sources are inferred simultaneously. Because the exact number of sources present in the data is not known a priori, this problem naturally requires a transdimensional inference scheme.

Among the several classes of sources expected to populate the LISA band, Galactic binaries (GB) are the most numerous, as the number of detectable signals is $\mathcal{O}(10^4)$, posing a major challenge in the analysis. 

We therefore apply the proposed algorithm to a highly downscaled version of this problem, in which one year of data contains the superposition of 10 Galactic binaries.

This experiment should be interpreted as a controlled proof of concept rather than as a realistic LISA global-fit analysis. The simplified setup allows us to test whether the proposed sampler can recover multiple overlapping Galactic binaries while inferring the number of sources and reusing posterior information from shorter observing durations.

Each binary is parametrized by the following set of parameters $\theta_j$:
\begin{equation}
  \theta_j =
  \{ f_0, \dot f, A, \beta, \lambda, \psi, \iota, \phi_0 \}_j ,
\end{equation}
where $f_0$ is the reference frequency, $\dot f$ the frequency derivative, $A$
the amplitude, $(\beta,\lambda)$ the sky location, $\psi$ the
polarization angle, $\iota$ the inclination, and $\phi_0$ the initial phase.

We choose to work in the two noise-orthogonal data channels A,E, in a noise-free approximation and a constant power spectral density.
The parameters of each binary are drawn uniformly within the allowed ranges, except for the initial frequencies $f_0$, which are fixed on an equally spaced grid in the range [0.9998, 1.0006] $\mathrm{mHz}$, such that the sources are clearly identifiable.
Given data $d$ observed in each channel $ c \in \{A,E\}$ and the corresponding projected template in such channels $h_c$,\footnote{We use \texttt{JAXGB} (\hyperlink{https://gitlab.com/lisamission/jaxgb}{https://gitlab.com/lisamission/jaxgb}) for computing GB signals.}
the log-likelihood in the frequency domain is:
\begin{align}
  \log \mathcal{L}(d\mid k,\boldsymbol{\theta})
  &\propto
  -\frac{1}{2}
  \sum_c
  \left(
    d_c-h_c(\boldsymbol{\theta})
    \mid
    d_c-h_c(\boldsymbol{\theta})
  \right)_c, 
  \label{eq:gw_likelihood}
\end{align}

where
\begin{equation}
  (a\mid b)_c
  =
  4\,\mathrm{Re}
  \sum_{f>0}
  \frac{a^*(f)b(f)}{S_c(f)}\,\Delta f .
\end{equation}

$\Delta f$ represents the frequency resolution, given by the inverse of the observation time, and $S_c(f)$ is the one-sided power spectral density (PSD) of the noise in each channel. Given the narrow frequency range, we choose a uniform PSD.
The optimal signal to noise ratio $\rho_\text{opt}$ for each source, computed as $\rho_\text{opt}^2 = \sum_c{(h_c, h_c)_c}$ lies between 10 and 40.

\begin{figure}[h!]
  \centering
  \resizebox{0.5\textwidth}{!}{
\includegraphics{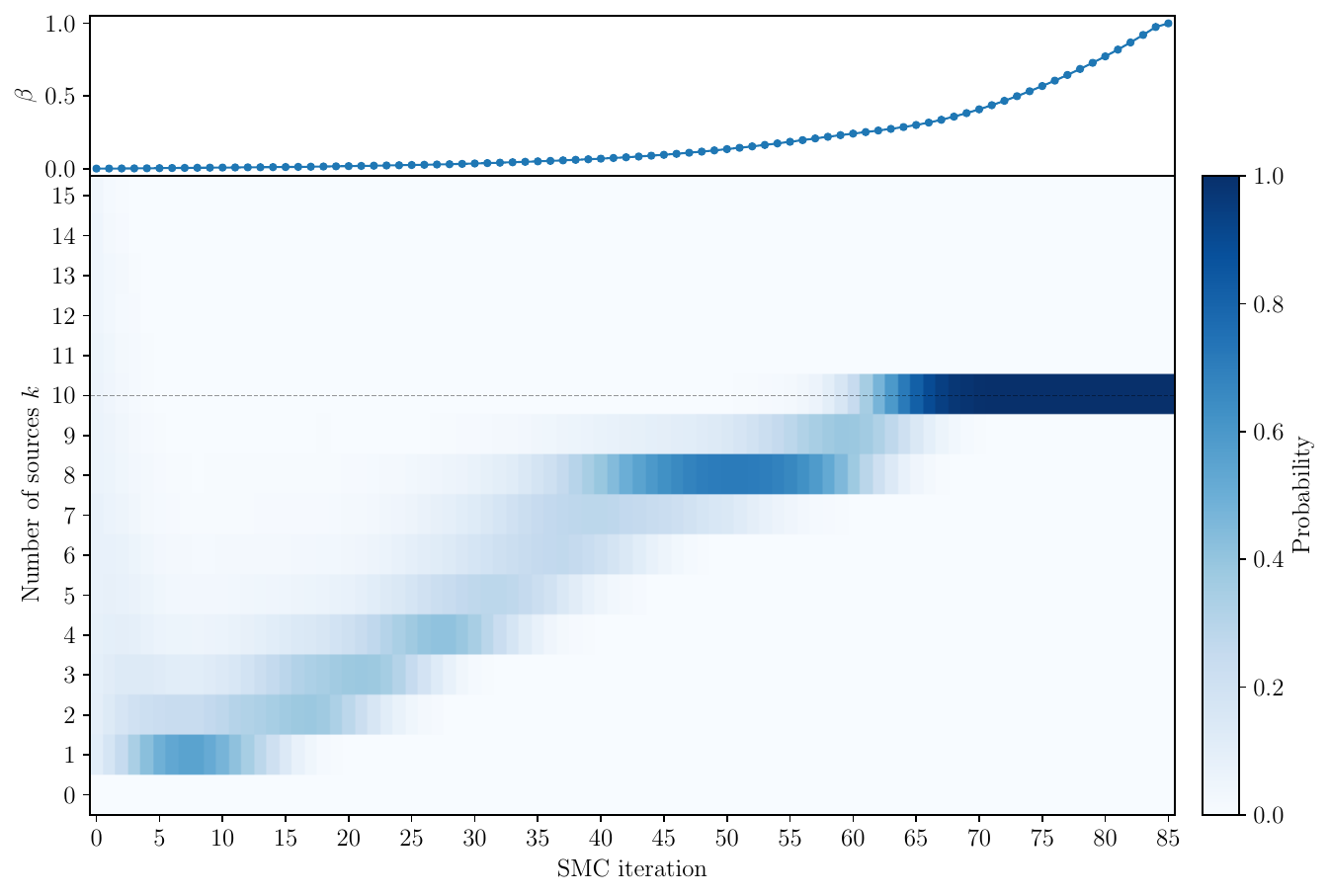
}}
  \caption{  \textit{Bottom panel:}
  Evolution of the posterior over $k$ as a function of the SMC iterations for the LISA Galactic Binaries problem (sec. \ref{sec:lisa_application}). For each iteration, we report the histogram on the number of components $k$ as a heatmap, such that a darker color indicates a higher probability.  
 \textit{Top panel:}
 Behaviour of the inverse temperature $\beta$ as a function of the SMC iterations.}
  \label{fig:heatmap_binaries}
\end{figure}

\begin{figure}[h!]
  \centering
  \resizebox{0.5\textwidth}{!}{
\includegraphics{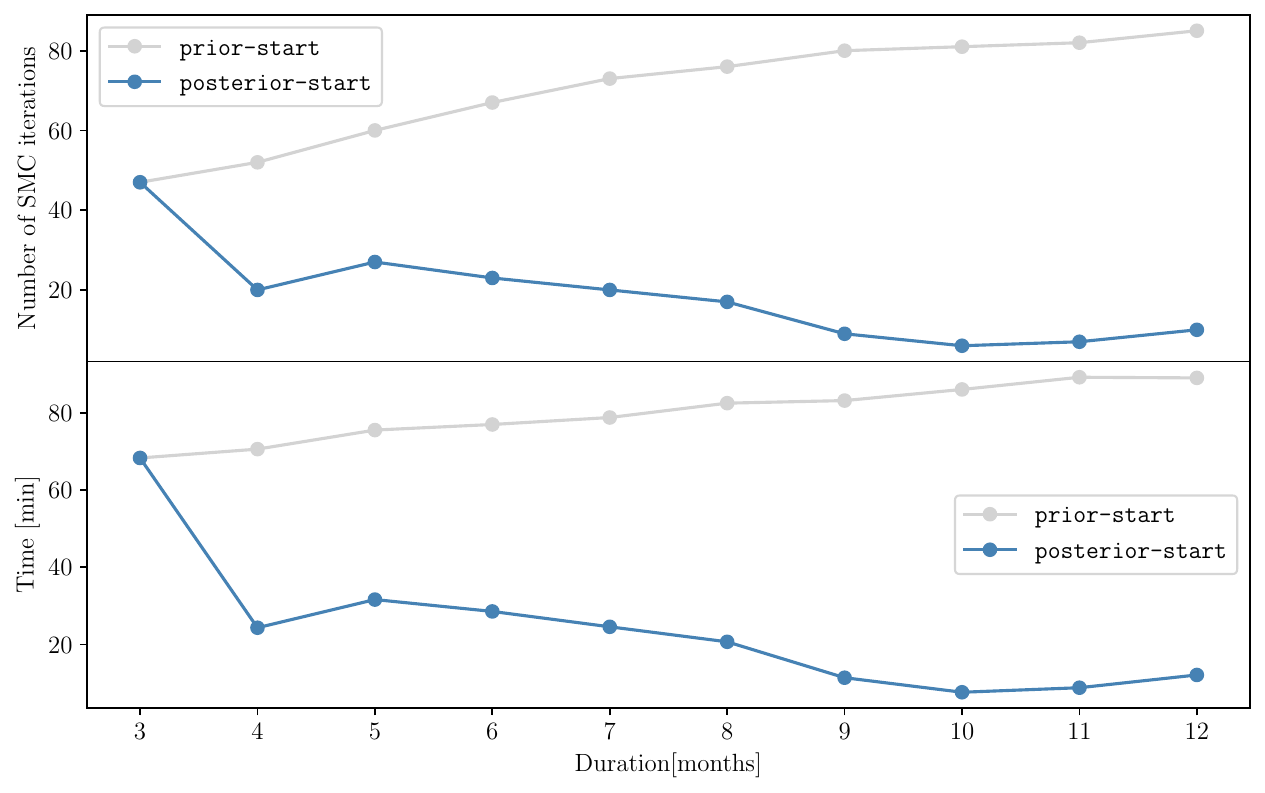
}}
  \caption{Comparison between the total number of SMC iterations (top panel) and of the runtime (bottom panel) in the \texttt{prior-start} and \texttt{posterior-start}  case as a function of the duration of the portion of the signal considered, for the LISA galactic binary case (sec. \ref{sec:lisa_application}). }
  \label{fig:gain_smc_iteration}
\end{figure}

\begin{figure}[h!]
  \centering
  \resizebox{0.5\textwidth}{!}{
\includegraphics{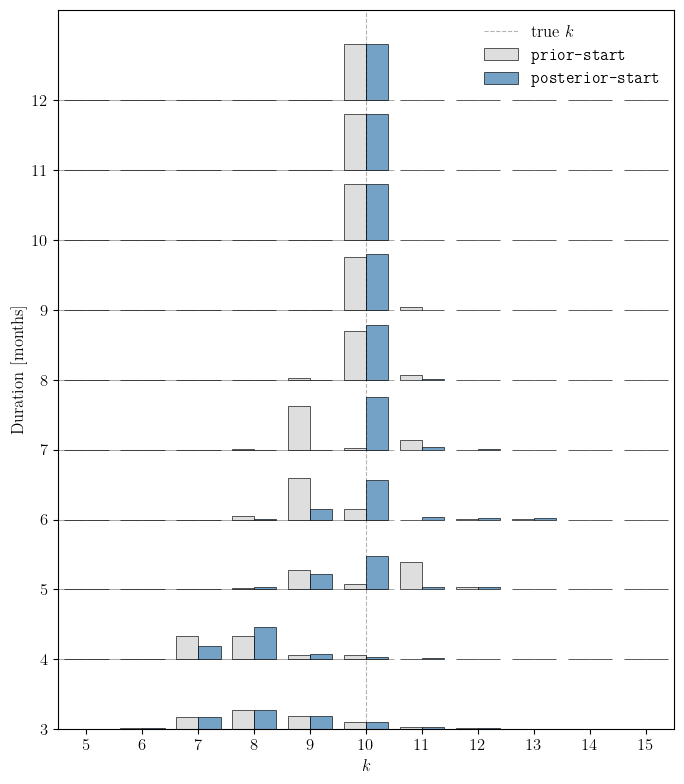
}}
  \caption{Comparison between the posterior over the number of galactic binaries found in data for the \texttt{prior-start} and \texttt{posterior-start}  case as a function of the duration of the portion of data considered. }
  \label{fig:k_hist_comparison_LISA}
\end{figure}

We run the algorithm with the same settings as the previous example, except for the number of particles that is increased to 40000 in order to deal with a larger and more complex parameter space.

We start by considering the full set of data, that has a duration of 1 year, running the algorithm in the \texttt{prior-start} case. 
In fig. \ref{fig:signal_recovery_lisa} we report the posterior over the inferred signal, as a shaded region with vertical hatches, while the dashed black line represents the injected value, demonstrating the faithful recovery of the injected signal. 
We also plot in fig. \ref{fig:heatmap_binaries} the behavior of the posterior over the number of galactic binaries found in data as a function of the SMC iteration, going from the prior to the posterior. 
We again find the same structure as the example before, with the posterior peaking at lower value than the injected one at earlier stages of the SMC as a consequence of the tempering, while it converges to the injected value in the last iterations. 
It may be also noticed that the posterior has a sharp peak at approximately $k = 8 $ around iteration 50, that corresponds to the loudest injected sources.

Switching to the \texttt{posterior-start} case we now have that, starting from a duration of three months, each set of data is longer than the previous one by 1 month, until the full 1 year of data, and we perform the analysis reusing the result obtained with the previous duration.
Again, for comparison, we run also the SMC in the \texttt{prior-start} mode.
Fig. \ref{fig:k_hist_comparison_LISA} shows a comparison, for each duration of the data considered, between the posterior over $k$ obtained in the two cases. 
Unlike the Gaussian-pulse example the posteriors do not fully agree for some durations, with the result obtained in the \texttt{prior-start} mode being off the injected value.

In fig. \ref{fig:gain_smc_iteration} we plot the comparison between the number of iterations (top panel) and the runtime (bottom panel) needed to converge to the target distribution from the base in the two cases. As expected, the \texttt{posterior-start} analysis requires substantially fewer SMC iterations than the \texttt{prior-start} analysis. For observing durations longer than nine months, the reduction reaches approximately a factor of five, leading to a comparable reduction in runtime.
\footnote{The runtimes reported here are obtained on an NVIDIA A100 GPU.}

\begin{figure}[h!]
  \centering
  \resizebox{0.5\textwidth}{!}{
\includegraphics{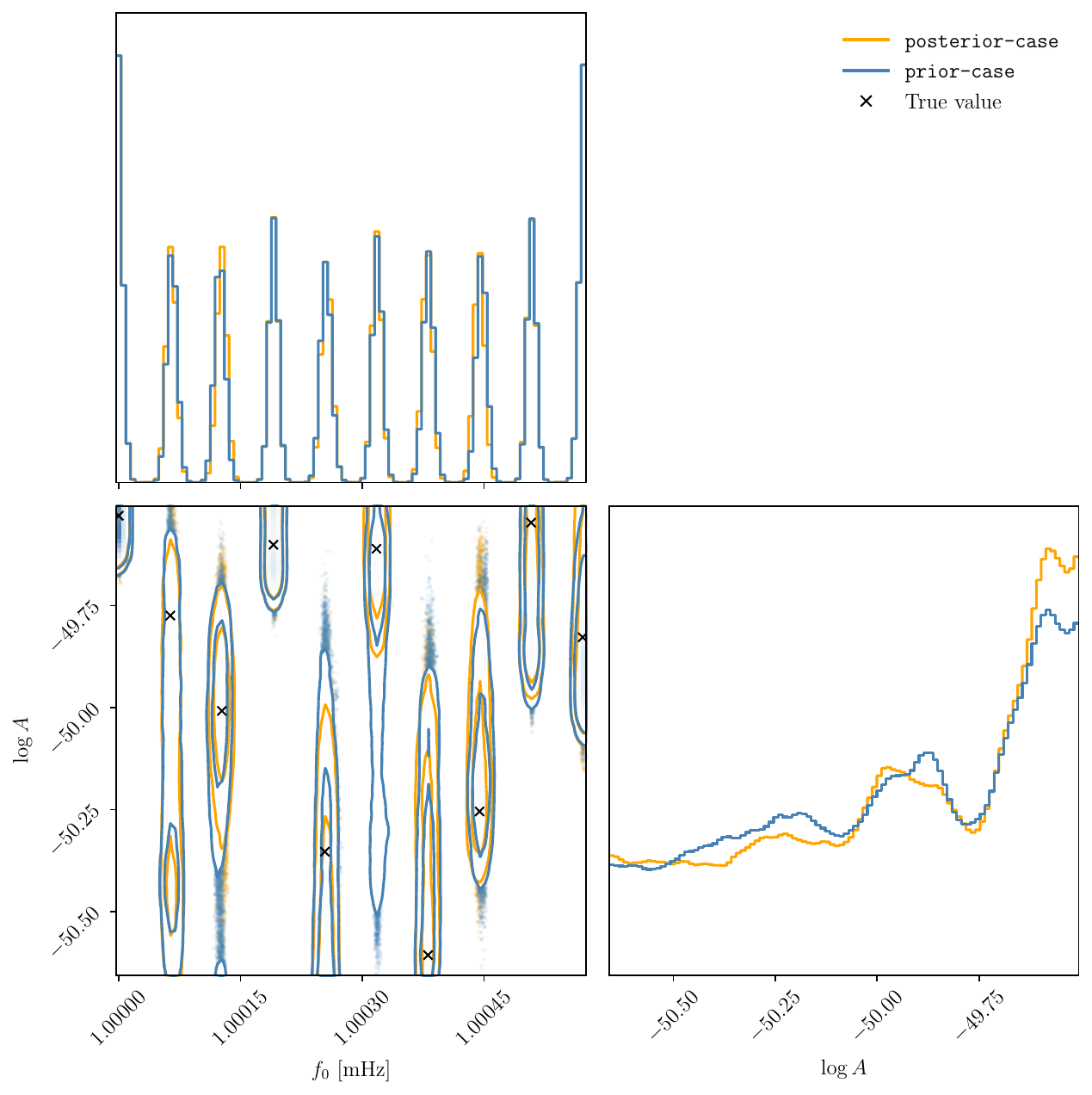
}}
  \caption{Comparison between posterior samples on the initial frequency $f_0$ and the logarithm of the amplitude $\log A$ of the galactic binaries found in data, obtained analyzing 12 months of data in the \texttt{prior-start} and \texttt{posterior-start}. 
  }
  \label{fig:corner_LISA}
\end{figure}

Finally, we want to compare the posterior samples obtained in the two cases when analyzing the full set of data. 
We do so by combining a subset of parameters, $f_0$ and $\log A$ , of each particle at the last iteration and plotting it in fig.\ref{fig:corner_LISA}. Except for a small portion of the parameters space, in which the support of the posterior in the \texttt{prior-start} case is larger, the results of the two methods show good agreement.

\section{Discussion}
\noindent

The proposed algorithm offers several advantages for both fixed-dimensional and trans-dimensional sampling:
\begin{itemize}
\item \underline{\textbf{ Parallelization}}\\ 
       As we have seen in the previous sections, each particle can be mutated independently. Therefore mutations, which are also the most computationally expensive, can be easily parallelized, distributing the computation across several CPUs or GPUs.
       In this work, we used a very specific transition kernel, nonetheless the above is valid for any trans-dimensional algorithm. 
       In short, the SMC scheme offers a clear method to easily parallelize a sampler, that would otherwise rely on the construction of a Markov chain, that is intrinsically serial.

\item \underline{\textbf{Automatic tempering scheme }}\\ 
      Tempering the posterior is a key ingredient for enhancing the exploration of the parameter space in the case of highly multimodal posterior distributions.

      Through the automatic choice of the temperature ladder, by requiring a constant ESS, SMC  offers a clear tempering scheme and a compelling alternative to parallel adaptive tempering.  
    The method shares some similarities with the adaptive parallel-tempered
RJMCMC currently used in Eryn \cite{Karnesis:2023ras}. In both cases, the
target distribution is explored through a tempered sequence of distributions
and reversible-jump moves are used to connect models with different numbers of
sources. However, the role of tempering is different. In parallel-tempered
RJMCMC, several Markov chains are evolved at fixed inverse temperatures and
information is exchanged through swap moves between neighbouring temperatures.
In the present approach, instead, the inverse temperature defines a sequence of
SMC targets: particles are reweighted, resampled, and then rejuvenated using a
mixture of fixed dimensional and trans-dimensional moves.
This distinction has two practical consequences. First, the SMC formulation is
naturally parallel over particles at each tempering level, with synchronization
required only for global operations such as ESS estimation, resampling, and the
adaptation of the next inverse temperature. Second, the SMC weights provide a
direct estimate of the normalizing constant, and hence of the Bayesian evidence, fundamental for model selection tasks, 
whereas in parallel-tempered RJMCMC this typically requires an additional
post-processing estimator, such as thermodynamic integration or related
methods.

\item \underline{\textbf{Update of results with new data}}\\
SMC offers a natural mechanism for updating posterior samples as new data become available. This not only improves the overall efficiency of the analysis, but it can be a substantial advantage in several situations. 

First, it can act as an additional form of tempering. Since the SNR of sources increase as data are accumulated, starting from a shorter data segment and then assimilating additional data can improve exploration of the parameter space, especially when the posterior to prior ratio is expected to be very small.

Second, it can play a crucial role in low-latency analysis, when rapid source-parameter estimates are essential, as in the case of massive black-hole binaries in LISA \cite{Magee:2021xdx, Li:2021mbo, Nobili:2026cnr}.
\end{itemize}
There are several aspects in which the proposed algorithm can be improved. 
First, despite the population nature of the algorithm and the very high efficiency of within-model moves with NUTS, using the prior distribution as the birth proposal can be extremely inefficient in full-scale scenarios.
Therefore, more efficient proposals should be used. 
For instance, since at every SMC step the particles approximate each tempered distribution, one could use such distribution as a proposal density\cite{Korsakova:2024sut}.
Moreover, the scaling of the method to full scenarios and hence higher dimensions may require a larger number of particles to faithfully approximate the distributions across SMC steps, creating a potential memory bottleneck on memory constrained hardware.

Additionally, in the current method the inclusion of new data doubles the likelihood evaluations, since the target distribution is eq. \eqref{eq: reuse_bridge}. A natural step forward would be to write eq. \eqref{eq: reuse_bridge} such that the ratio simplifies, although we expect to be non-trival given the generally correlated nature of noise.
Alternatively, a density estimator can be used to approximate the previous distribution\cite{Williams:2025aar}.
These investigations are left for future work.

\section{Conclusions}\label{sec:Conclusions}
We have presented a trans-dimensional Sequential Monte Carlo algorithm that uses the No-U-Turn Sampler for efficient within-model exploration and reversible-jump moves to transition between models with different numbers of components,  while also being able to incorporate information from results obtained with a shorter period of data.

We first applied the method to a toy problem, in which we inferred the number of gaussian pulses in data. 
We then validated the algorithm on a highly simplified LISA scenario, in which we injected 10 galactic binaries in a narrow frequency region, demonstrating its ability to correctly recover both the number of injected signals and their parameters. 
The application of the algorithm to progressively more realistic problems is left for future work.

In the examples considered, \texttt{posterior-start} analyses produce final posterior estimates consistent with \texttt{prior-start} analyses while requiring fewer SMC iterations and shorter runtime,  up to a factor of five.

These results provide an initial proof-of-concept that trans-dimensional SMC with gradient-based within-model mutations can be a viable route toward scalable LISA global-fit inference, since it delivers an extremely parallel and efficient exploration of the parameter space in high-dimensional cases. Nonetheless, it can be useful in several other fields in gravitational-wave astronomy that require parallel and efficient trans-dimensional inference, such as glitch-robust parameter estimation and binary black-hole population analyses.

\subsection*{Acknowledgments}
We thank Michael Williams for carefully reading the manuscript and for providing useful comments.
We acknowledge ISCRA for awarding this project access to the LEONARDO supercomputer, owned by the EuroHPC Joint Undertaking, hosted by CINECA (Italy). GD acknowledges financial support from the National Recovery and
Resilience Plan (PNRR), Mission 4 Component 2 Investment 1.4, National Center for HPC, Big Data and
Quantum Computing, funded by the European Union, NextGenerationEU, CUP B83C22002830001.

\appendix

\printbibliography

\end{document}